\documentclass[a4paper,onecolumn,11pt,accepted=2025-08-27]{quantumarticle}
\pdfoutput=1

\usepackage{thmtools}
\usepackage{thm-restate}
\usepackage{array, booktabs}
\usepackage[utf8]{inputenc}
\usepackage[english]{babel}
\usepackage[T1]{fontenc}
\usepackage{amsmath}
\usepackage{amssymb}
\usepackage{hyperref}
\usepackage{algorithm}
\usepackage{algpseudocode}
\usepackage{multirow}
\usepackage{makecell}
\usepackage{tikz}
\usepackage{quantikz}
\usepackage{lipsum}

\newtheorem{theorem}{Theorem}
\newtheorem{lemma}{Lemma}
\newtheorem{definition}{Definition}

\newtheorem{Cor}{Corollary}

\usepackage[numbers,sort&compress]{natbib}
\begin{document}

\title{Adaptivity is not helpful for Pauli channel learning}
% \title{Adaptivity Is Not Helpful for Pauli Channel Learning.}
\author{Xuan Du Trinh}
\email{xtrinh@cs.stonybrook.edu}
% \author{Lídia del Rio}
\affiliation{Department of Computer Science, Stony Brook University, Stony Brook, NY 11794, USA}
\orcid{0009-0009-5610-462X}

\author{Nengkun Yu}
\email{nengkun.yu@cs.stonybrook.edu}
\affiliation{Department of Computer Science, Stony Brook University, Stony Brook, NY 11794, USA}
\orcid{0000-0003-1188-3032}

% \date{April 18, 2025}

\maketitle

\begin{abstract}
We prove that adaptive strategies offer no advantage over non-adaptive ones for learning and testing Pauli channels using entangled inputs. This key observation allows us to characterize the query complexity for several fundamental tasks by translating optimal classical estimation algorithms into the quantum setting. First, we determine the tight query complexity for learning a Pauli channel under the general $\ell_p$ norm, providing results that improve upon or match the best-known bounds for the $\ell_1, \ell_2,$ and $\ell_\infty$ distances. Second, we resolve the complexity of testing whether a Pauli channel is a white noise source. Finally, we show that the optimal query complexities for estimating the Shannon entropy and support size of the channel's error distribution, and for estimating the diamond distance between two Pauli channels, are all $\Theta\left(\tfrac{4^n}{n\epsilon^2}\right)$.
\end{abstract}

\section{Introduction}
Quantum noise is a fundamental concept in quantum physics, challenging our ability to demonstrate and exploit the advantages of quantum information processing. In quantum cryptography, noise primarily influences quantum communication tasks by generating errors, thereby reducing the secret key rate of quantum key distribution protocols. To build large-scale and fault-tolerant quantum computers, mitigating quantum noise and performing error correction are essential. For these reasons, modeling and evaluating quantum noise always remain a topic of interest.

In the context of quantum error correction, bit flips and phase flips represent the most fundamental types of errors that illustrate the impact of noise on qubit systems. Consequently, the Pauli channel is widely adopted as a standard noise model.

% \begin{definition} [Pauli Channel]
%     Let $\rho$ denote the density operator of an $n$-qubit system, the Pauli channel $\mathcal{P}$ is a completely positive and trace-preserving (CPTP) map: 
% \begin{equation}
%     \rho \longmapsto \sum_{i\in\{0,1,2,3\}^n}P(i)\cdot\tau_{i}\rho\tau_{i}^\dagger
% \end{equation}
% where $\tau_i=\sigma_{i_1}\otimes\sigma_{i_2}\otimes...\otimes\sigma_{i_n}$ is a tensor product of Pauli operators from the set $\{\sigma_0$, $\sigma_1$, $\sigma_2$, $\sigma_3\}$ of Pauli matrices and $P(i)$ represents the probability of the error denoted by $\tau_i$.
% \end{definition}

% \begin{definition}[Pauli Channel]
%     Let $\rho$ denote the density operator of an $n$-qubit system.  
%     The Pauli channel $\mathcal{P}$ is a completely positive and trace-preserving (CPTP) map:
%     \begin{equation}
%         \rho \longmapsto \sum_{i\in\{0,1,2,3\}^n} P(i)\cdot \tau_{i}\rho\tau_{i}^\dagger ,
%     \end{equation}
%     where $\tau_i = \sigma_{i_1}\otimes\sigma_{i_2}\otimes \cdots \otimes\sigma_{i_n}$ 
%     is a tensor product of Pauli operators taken from the set
%     \[
%         \sigma_0 = I = \begin{bmatrix}1 & 0 \\ 0 & 1\end{bmatrix}, \quad
%         \sigma_1 = X = \begin{bmatrix}0 & 1 \\ 1 & 0\end{bmatrix}, \quad
%         \sigma_2 = Y = \begin{bmatrix}0 & -i \\ i & 0\end{bmatrix}, \quad
%         \sigma_3 = Z = \begin{bmatrix}1 & 0 \\ 0 & -1\end{bmatrix},
%     \]
%     and $P(i)$ represents the probability of the error denoted by $\tau_i$.
% \end{definition}

\begin{definition}[Pauli Channel]
For an $n$-qubit system with density operator $\rho$, the Pauli channel $\mathcal{P}$ is the completely positive and trace-preserving (CPTP) map
\begin{equation}
    \rho \longmapsto \sum_{i\in\{0,1,2,3\}^n} P(i)\,\tau_i \rho \tau_i^\dagger ,
\end{equation}
where $\tau_i = \sigma_{i_1} \otimes \sigma_{i_2} \otimes \cdots \otimes \sigma_{i_n}$
is a tensor product of Pauli operators chosen from
\[
\begin{aligned}
\sigma_0 &= I = \begin{bmatrix}1 & 0 \\ 0 & 1\end{bmatrix}, &
\sigma_1 &= X = \begin{bmatrix}0 & 1 \\ 1 & 0\end{bmatrix}, \\
\sigma_2 &= Y = \begin{bmatrix}0 & -i \\ i & 0\end{bmatrix}, &
\sigma_3 &= Z = \begin{bmatrix}1 & 0 \\ 0 & -1\end{bmatrix},
\end{aligned}
\]
and $P(i)$ is the probability associated with the error operator $\tau_i$.
\end{definition}

The probability distribution $P$ characterizes the channel $\mathcal{P}$. 
When a quantum state $\rho$ passes through a noisy device modeled by the Pauli channel, 
information about $P$ can be obtained from measurements on the output state $\mathcal{P}(\rho)$. 
By preparing known input states and selecting appropriate measurement settings, either projective or generalized, one can collect samples from $P$.

%of size $N$ after repeating the experiment $N$ times. 

One might consider a specific parameterization for $P$ while analyzing data samples based on physical models. In particular,  several well-studied single-parameter Pauli channels have been analyzed in the literature, such as the bit flip channel, the dephasing channel and the depolarizing channel \cite{NoiseSourceModeling}. However, our analysis treats $P$ in full generality. Learning and testing a probability distribution without presupposing its form presents a challenging and contemporary problem in statistics and learning theory. This paper explores the complexity of various related tasks, including learning the Pauli channel and testing some aspects of this model.

% \begin{definition}[Learning Complexity]
% For $\epsilon,\hspace{1mm} \delta \in (0,1)$, we define the learning complexity as the number of times we query the Pauli channel to output an estimation $\hat{P}$ that is $\epsilon$-close to $P$, with respect to a predefined channel distance and an error probability at most $\delta$.    
% \end{definition}

\begin{definition}[Learning Complexity]
For $\epsilon, \delta \in (0,1)$, the learning complexity is defined as the number of queries to the Pauli channel required to produce an estimate $\hat{P}$ that is $\epsilon$-close to $P$ with respect to a predefined channel distance, with error probability at most $\delta$. 
\end{definition}

 % To simplify the analysis, in this study, if the value of the soundness $\delta$ is not specified, we understand that it is fixed as a constant, namely $\delta=1/3$. In literature, the learning complexity has been investigated in both senses: the number of the Pauli channel applications and the number of measurements. In this paper, we find that distinguishing between these two notions is unnecessary. 

To simplify the analysis, when the soundness parameter $\delta$ is not specified, it is taken to be a constant, namely $\delta = 1/3$. In the literature, learning complexity has been investigated in two senses: either in terms of the number of Pauli channel applications or the number of measurements. In this paper, we find that distinguishing between these two notions is unnecessary.

% Having already established the definition of the learning complexity, we need a scheme to discriminate between the actual parameters of the quantum channel and the distribution learned from data samples. 
% \begin{definition}[Channel Distance Metrics]
% The distance $\ell_p$, with $p\geq 1$, between two probability distributions $P$ and $\hat{P}$ is defined by
% \begin{equation}
%     \ell_p(\hat{P}, P)=||\hat{P}-P||_p=\left (\sum_{i\in \{0,1,2,3\}^n} |\hat{P}(i)-P(i)|^p\right )^{\frac{1}{p}}.
% \end{equation}    
% \end{definition}

% It is noteworthy that $\ell_1(\hat{P}, P)=\sum_{i}|\hat{P}(i)-P(i)|$ represents twice the total variation distance, while $\ell_{\infty}(\hat{P}, P)=\max_{i}{|\hat{P}(i)-P(i)|}$ signifies the maximal difference between two distributions at a single coordinate, and $\ell_2(\hat{P}, P)=\sqrt{ \sum_{i} |\hat{P}(i)-P(i)|^2}$ corresponds to the common Euclidean distance.

Having defined learning complexity, it is natural to introduce a metric to compare the true parameters of the quantum channel with the distribution learned from data samples. 

\begin{definition}[Channel Distance Metrics]
The distance $\ell_p$, with $p \geq 1$, between two probability distributions $P$ and $\hat{P}$ is given by
\begin{equation}
    \ell_p(\hat{P}, P) = \|\hat{P} - P\|_p =
    \left( \sum_{i \in \{0,1,2,3\}^n} |\hat{P}(i) - P(i)|^p \right)^{1/p}.
\end{equation}    
\end{definition}

In particular, $\ell_1(\hat{P}, P) = \sum_i |\hat{P}(i)-P(i)|$ equals twice the total variation distance. The metric $\ell_\infty(\hat{P}, P) = \max_i |\hat{P}(i)-P(i)|$ measures the maximum pointwise difference between the two distributions. Finally, $\ell_2(\hat{P}, P) = \sqrt{\sum_i |\hat{P}(i)-P(i)|^2}$ corresponds to the standard Euclidean distance.

Previous research has yielded remarkable results in estimating Pauli channels under various conditions, including scenarios with and without entanglement and employing different metrics for evaluation (see Table \ref{tab: compare learning bounds}). In the entanglement-free scheme, Flammia and Wallman introduced an algorithm requiring $O(n2^n/\epsilon^{2})$ measurements to estimate Pauli channels in norm $\ell_2$ in \cite{flammia2020paulichannel}. Recently, Fawzi et \textit{al.} refined this algorithm and established an upper bound of $O(n2^{3n}/\epsilon^{2})$ measurements in the context of norm $\ell_1$  \cite{fawzi2023lowerbound_paulichannel}. While focusing on the $\ell_\infty$ norm and without-entanglement condition, Flammia and O’Donnell related learning of Pauli channels to the Population Recovery problem. They provided an algorithm that uses the channel $O(1/\epsilon^{2})\log(n/\epsilon)$ times \cite{flammia2021pauli_populationrecovery}. In another study in the same context, Chen et \textit{al.} demonstrated that $\Theta(2^n/\epsilon^{2})$ measurements are required for learning each eigenvalue $P(i)$ \cite{LiangJiang2023tightbound}. Earlier, Chen et al. showed that entanglement provides an exponential advantage, with an algorithm that applies the channel $O(n/\epsilon^{2})$ times \cite{LiangJiang2022bound_entanglement_advantage}, leading to an upper bound of $O(4^n/\epsilon^2)$ in the $\ell_1$ distance via the Parseval–Plancherel identity \cite{fawzi2023lowerbound_paulichannel}. In another results, Flammia and O'Donnell highlighted that $\Theta(1/\epsilon^{2})$ samples of $P$ are both necessary and sufficient for classically learning $P$ in the $\ell_\infty$ norm. They further stated that Bell measurements can obtain these samples while ancilla, i.e., entanglement, is available \cite{flammia2021pauli_populationrecovery}. These studies, alongside others \cite{Noise-tailoring-Wallman-PhysRevA.94.052325,Chen-LiangJiang-learnability-Pauli-NatureComm,Rouze2023-França}, highlight the complexity and diversity of techniques in quantum channel discrimination. Motivated by this diversity, our paper aims to establish tight bounds for the learning problem across any $\ell_p$ norm, where $p\geq 1$.

{
\renewcommand{\arraystretch}{1.5} % Adjusts the height of each row
\setlength{\tabcolsep}{11pt} % Adjusts the space between columns
\begin{table}[!ht]
\centering
\begin{tabular}{|c|c|c|c|}
\hline
\textbf{p} & \textbf{This study} & \textbf{With entanglement} & \textbf{Without entanglement} \\
\hline
\(p=1\) & \( \Theta(4^n/\epsilon^2) \) & \( \Tilde{O}(4^n/\epsilon^{2})  \)\cite{LiangJiang2022bound_entanglement_advantage,fawzi2023lowerbound_paulichannel} & \( O(n2^{3n}/\epsilon^{2})\cite{flammia2020paulichannel}\) \\
\hline
\(p=2\) & \( \Theta(1/\epsilon^2) \) & \( None \) & \( O(n2^{n}/\epsilon^{2})\cite{flammia2020paulichannel} \) \\
\hline
\( p = \infty \) & \( \Theta(1/\epsilon^2 ) \) & \( \Theta(1/\epsilon^2) \cite{flammia2021pauli_populationrecovery}\footnotemark[1] \) & \( \Omega(1/\epsilon^2), \hspace{1mm}O(1/\epsilon^2)\log(n/\epsilon)\)\cite{flammia2021pauli_populationrecovery} \\
\hline
\end{tabular}
\caption{Comparison of learning complexities obtained in Corollary \ref{corollary tight bound for learning} with existing sample complexity bounds in literature for different common values of $p$. The two last columns list the results previously obtained with and without entanglement resources in terms of the number of measurement rounds or the number of Pauli channel applications.}
\label{tab: compare learning bounds}
\end{table}
\footnotetext[1]{Flammia and O'Donnell connected the problem of learning $\mathcal{P}$ to the estimation of the discrete distribution $P$ in \cite{flammia2021pauli_populationrecovery}, citing the classical sample complexity of $\Theta(1/\epsilon^{2})$ for learning $P$ as reported in \cite{Canonne2020}. However, they did not provide a proven lower bound for the quantum query complexity.}
}

In addition to quantum channel discrimination, adaptive quantum learning algorithms play a significant role. These refer to the strategy of dynamically adjusting the choice of quantum measurements and input states based on the outcomes of previous measurements (Figure \ref{fig: adaptive strategies}). This approach contrasts with non-adaptive strategies, where the measurement bases are predetermined before experimentation. Notably, the algorithms in \cite{flammia2020paulichannel} employ adaptivity as a crucial component. Subsequently, Fawzi et \textit{al.} in their recent work explored lower bounds for adaptive and non-adaptive strategies in the context of $\ell_1$ and $\ell_\infty$ and specifically in scenarios devoid of entanglement resources \cite{fawzi2023lowerbound_paulichannel}. Their results show gaps between the aforementioned lower and upper bounds. This raises an essential question regarding the actual efficacy of adaptivity.

\begin{figure}[ht]
    \centering
    \begin{quantikz}
    \lstick{\ket{0^{k}}} & \gate[wires=2]{A} & \qw  &  \gate[wires=2]{A}   & \qw  & \textit{.......}&\gate[wires=2]{A} & \qw   &  \gate[wires=2]{A}& \qw\\
    \lstick{\ket{0^n}} & \qw   & \gate{\mathcal{P}} &  \qw   & \gate{\mathcal{P}}  & \textit{.......}& \textit{.......}  & \gate{\mathcal{P}} &  \qw & \\
\end{quantikz}
    \caption{Adaptive strategy for the sampling task: The main $n$-qubit system and a possible ancilla of $k$ qubits are initialized respectively in the state $\ket{0^n}$ and $\ket{0^{k}}$. The operation $A$ denotes an adaptive procedure which may involve quantum operations and measurements whose settings can be chosen based on previous outcomes. The gate $\mathcal{P}$ processes the main system in order to extract information from $P$. The circuit outputs a sample from $P$.}
    \label{fig: adaptive strategies}
\end{figure}
We present the main theorems of our study, addressing the role of adaptivity in the learning process of the Pauli channel when entanglement resources and ancillary registers are accessible. These theorems enable further analysis of other aspects of Pauli channels.

\begin{theorem}
Given a Pauli channel $\mathcal{P}$, the associated random variable $P$, and the capability to prepare maximally entangled two-qubit states. It is established that a bisimulation relationship exists between the Pauli channel $\mathcal{P}$ and the sampling process of its associated probability distribution P. Specifically, the following statements hold
\begin{itemize}
\item The Pauli channel $\mathcal{P}$ can be simulated utilizing the random variable $P$.
\item Obtaining one individual sample from $P$ necessitates exactly one application of the channel $\mathcal{P}$.
\end{itemize}
\label{theorem: main bisimulation}
\end{theorem}

This theorem gives an insight that, in terms of information, accessing the channel $\mathcal{P}$ is equivalent to accessing the random variable $P$. The intuition behind the first point is straightforward since $P$ effectively characterizes $\mathcal{P}$, and the second point will be demonstrated by introducing a sampling algorithm. The equivalence signifies that tasks such as learning or estimating properties of $\mathcal{P}$ can be efficiently executed by sampling from the random variable $P$ and then applying a classical processing algorithm. The impossibility of drawing a larger number of samples $N$ than the number of channel applications $M$ (i.e., $N>M$) is a key element in the proof of Theorem \ref{theorem: adaptivity is not useful}. The important implication here is that if the sample complexity for a task is $\mathcal{C}$, then the query complexity on $\mathcal{P}$ directly aligns with $\mathcal{C}$. For instance, achieving an $\epsilon$-close approximation in learning the distribution $P$ within the $\ell_p$ norm requires $\Theta(poly(\frac{1}{\epsilon},2^n))$ samples, then an equal number of queries to the channel $\mathcal{P}$ is required. The upper and lower bounds are all translated from sample complexity to query complexity. A direct and powerful consequence of this one-to-one correspondence between channel applications and samples is that adaptive strategies offer no advantage, a result we formalize in the following theorem.

\begin{theorem}
    In estimating any properties of channel $\mathcal{P}$ using samples drawn from distribution $P$, adaptivity does not help to improve channel query complexity compared to non-adaptive strategies.
\label{theorem: adaptivity is not useful}
\end{theorem}

Theorem \ref{theorem: adaptivity is not useful} implies that there exists a non-adaptive algorithm that achieves optimal query complexity for the tasks that use samples from $P$. There will not be an improvement if one uses any adaptive strategy. The proofs of Theorem \ref{theorem: main bisimulation} and Theorem \ref{theorem: adaptivity is not useful} are presented in Section 2. The main technique used is an algorithm for the sampling of Pauli distribution $P$, that transfers one channel application to yield one sample of $P$. From these results, subsequent sections will discuss its applications in learning and testing problems, providing Corollaries \ref{corollary tight bound for learning}, \ref{corollary testing white noise}, \ref{corollary Estimate the noise level} and \ref{corollary estimate diamond} while referencing earlier groundbreaking studies \cite{waggoner2015lpdistance, VV11,VV17}. Corollaries \ref{corollary Estimate the noise level}  and \ref{corollary estimate diamond} will be more formally restated in Corollaries \ref{corollary Estimate the noise level formally} and \ref{Cor estimate diamond formally} while the characterization of Pauli channel's noise level and the diamond distance between two Pauli channels will be rigorously defined in the corresponding discussions in Section 3.

\begin{Cor} [Learning Complexity of Pauli Channels]
Let $\epsilon \in (0,1)$ and let $\mathcal{P}$ be an $n$-qubit Pauli channel. To learn $\mathcal{P}$ within $\epsilon$ with respect to the $\ell_p$ distance and with constant success probability, the number $N$ of required channel applications scales, up to $p$-dependent constant factors, as
   \begin{equation}
    N=\begin{cases}
        \Theta(4^{\frac{n(2-p)}{p}}\epsilon^{-2}) &\textit{ for } 1\leq p \leq 2 \textit{ and } \epsilon \leq 4^{-\frac{n(p-1)}{p}}\\
        \Theta(4^n) &\textit{ for } 1\leq p \leq 2 \textit{ and }  4^{-\frac{n(p-1)}{p}} \leq \epsilon \leq 2\cdot 4^{-\frac{n(p-1)}{p}}\\
        \Theta(\left(\frac{1}{\epsilon}\right)^{\frac{p}{p-1}}) \hspace{1.1cm} &\textit{ for } 1\leq p \leq 2  \textit{ and } \epsilon \geq 2\cdot 4^{-\frac{n(p-1)}{p}}\\
        \Theta(\epsilon^{-2})  &\textit{ for } 2\leq p\leq \infty
    \end{cases}
    \end{equation}
\label{corollary tight bound for learning}
\end{Cor}

\begin{Cor}[Complexity of White Noise Testing]
    Let $\epsilon \in (0,1)$ and let $P$ denote the discrete distribution characterizing an $n$-qubit Pauli channel $\mathcal{P}$. Up to constant factors, Table \ref{tab:Uniformity test bounds} shows the necessary and sufficient number of Pauli channel applications required to distinguish between the following two hypotheses with constant success probability:
\[
H_0: P = U_{4^n} \quad \text{and}\quad H_1: \ell_p(P, U_{4^n}) > \epsilon
\]
where $U_{4^n}$ is the uniform distribution over $4^n$ errors.
    \label{corollary testing white noise}
\end{Cor}

% \begin{Cor}[Noise Level Testing]
%     To estimate the noise level of the Pauli channel $\mathcal{P}$ characterized by the Shannon entropy of $P$ and number of Pauli errors that happen with non-zero probabilities, the necessary and sufficient number of Pauli channel applications is $\Theta(\frac{4^n}{n\epsilon^2})$.
%     \label{corollary Estimate the noise level}
% \end{Cor}

\begin{Cor}[Complexity of Noise Level Estimation]
Let $\mathcal{P}$ be a Pauli channel characterized by the distribution $P$. 
To estimate the noise level of $\mathcal{P}$ in terms of the Shannon entropy of $P$ and the support size of $P$ (the number of Pauli errors occurring with non-zero probability), 
the necessary and sufficient number of Pauli channel applications is 
$\Theta\!\left(\tfrac{4^n}{n\epsilon^2}\right)$.
\label{corollary Estimate the noise level}
\end{Cor}

\begin{Cor}[Complexity of Diamond Distance Estimation]
To estimate the diamond distance between two Pauli channels $\mathcal{P}_1$ and $\mathcal{P}_2$, the necessary and sufficient number of applications of each channel is $\Theta(\tfrac{4^n}{n\epsilon^2})$.
    \label{corollary estimate diamond}
\end{Cor}

% \begin{Cor}[Complexity of Diamond Distance Estimation]
% The necessary and sufficient number of applications of each Pauli channel $\mathcal{P}_1$ and $\mathcal{P}_2$ required to estimate their diamond distance within $\epsilon$ is 
% \[
% \Theta\!\left(\tfrac{4^n}{n\epsilon^2}\right).
% \]
% \label{corollary estimate diamond}
% \end{Cor}

{
\renewcommand{\arraystretch}{1.5} % Adjusts the height of each row
\setlength{\tabcolsep}{20pt} % Adjusts the space between columns
\begin{table}[!h]
\centering
\begin{tabular}{|c|c|c|c|}
\hline
\textbf{p value} & \textbf{Regime} & \textbf{Necessary} & \textbf{Sufficient} \\
\hline
\multirow{2}{*}{\(1 \leq p \leq 2\)} & \(   \epsilon \geq 2^{-\frac{2n(p-1)}{p}} \) & \( \epsilon^{-\frac{p}{2(p-1)}} \) & \( \epsilon^{-\frac{p}{2(p-1)}}\) \\
\cline{2-4}
& \( \epsilon < 2^{-\frac{2n(p-1)}{p}} \) & \( 2^{\frac{n(4-3p)}{p}}\epsilon^{-2} \) & \( 2^{\frac{n(4-3p)}{p}}\epsilon^{-2} \) \\
\hline
\multirow{3}{*}{\(2 < p < \infty\)} & \( \Theta\left( \frac{n}{4^n} \right) \geq \epsilon \) & \( \frac{n}{4^n\epsilon^2} \) & \( \frac{1}{2^n\epsilon^2} \) \\
\cline{2-4}
& \( \Theta\left( \frac{n}{4^n} \right) < \epsilon \leq \frac{1}{2^n} \) & \( \frac{1}{\epsilon} \) & \( \frac{1}{2^n\epsilon^2} \) \\
\cline{2-4}
& \( \epsilon > \frac{1}{2^n} \) & \( \frac{1}{\epsilon} \) & \( \frac{1}{\epsilon} \) \\
\hline
\multirow{2}{*}{\( p = \infty \)} & \( \Theta\left( \frac{n}{4^n} \right) \geq \epsilon \) & \( \frac{n}{4^n\epsilon^2} \) & \( \frac{n}{4^n\epsilon^2} \) \\
\cline{2-4}
& \( \Theta\left( \frac{n}{4^n} \right) < \epsilon \) & \( \frac{1}{\epsilon} \) & \( \frac{1}{\epsilon} \) \\
\hline
\end{tabular}
\caption{White noise testing query complexity bounds. The bounds indicate the necessary and sufficient number of channel queries to distinguish Pauli error distribution $P$ from the uniform distribution $U_{4^n}$, with respect to the $\ell_p$-distance $\epsilon$.}
\label{tab:Uniformity test bounds}
\end{table}
}

\section{Sampling with Bell State Measurement as An Optimal Strategy}
\subsection{Sampling Algorithm}
In the theory of quantum operations, Choi's isomorphism offers a representation of quantum channels. In this representation, the channel $\mathcal{P}$ acting on the $n$-qubit system $A$ is fully characterized by the quantum state
\begin{equation}
    \Lambda=(\mathbb{I} \otimes \mathcal{P})(\ket{\Phi_n}\bra{\Phi_n})
\end{equation}
where $\ket{\Phi_n}=\frac{1}{\sqrt{2^n}}\sum_{i=0}^{2^n-1}\ket{ii}$ is the maximally entangled state in the extension $A\otimes A^\prime$ with $A^\prime$ isomorphic to $A$. The fact that $\Lambda$ encodes all information about $\mathcal{P}$ implies that extracting information from $\Lambda$ is equivalent to ``measuring'' the quantum channel. It suggests a method for detecting Pauli errors using entangled states. Nevertheless, entangled measurement on the total system $A\otimes A^\prime$ is generally resource-intensive. So, we will employ $n$ Bell measurements on $n$ pairs of qubits. This strategy was introduced by \cite{Hangleiter2024}, where the authors demonstrated its application in extracting information and identifying errors in quantum circuits. Our main contribution lies in the discovery of Theorems \ref{theorem: main bisimulation} and \ref{theorem: adaptivity is not useful} with their implications, where the strategy is specifically applied to Pauli channels. We adapt our notations and conventions as described below.

A simple example will give an intuition of the approach. We start with the preparation of a pair of qubits in a maximally entangled state $\ket{\Phi^+} = \frac{1}{\sqrt{2}}(\ket{00} + \ket{11})$. One of the two qubits is then subjected to a Pauli channel $\mathcal{P}$. Subsequently, a Bell measurement is performed on the qubits (Figure \ref{fig: Bell measurement detecting error}).
\begin{figure}[!h]
    \centering
    \begin{quantikz}
    \lstick{$\ket{0}$} & \gate{H} & \ctrl{1} & \qw & \ctrl{1} & \gate{H} & \meter{} & \qw \\
    \lstick{$\ket{0}$} & \qw      & \targ{}  & \gate{\mathcal{P}} & \targ{} & \qw & \meter{} & \qw \\
\end{quantikz}
    \caption{The first Hadamard and CNOT gates create a maximally entangled state. Then, the second qubit is subject to a single-qubit Pauli channel. The second CNOT and Hadamard gates help to project Bell measurement to the measurement in the computational basis.}
    \label{fig: Bell measurement detecting error}
\end{figure}\\
\textbf{Effect of Pauli Errors:} If a Pauli error occurs on one qubit among an EPR pair, it would transform the initial entangled state \(\ket{\Phi^+}\) into another Bell state. To summarize, the measurement outcomes in a Bell measurement provide information about the types of Pauli errors that may have affected the qubits:

\begin{itemize}
    \item $\ket{\sigma_0}=\ket{\Phi^+} = \frac{1}{\sqrt{2}} (\ket{00} + \ket{11})=(\mathbb{I}_2\otimes \sigma_0)\ket{\Phi^+}$ indicates no errors ($\sigma_0=\mathbb{I}_2$).
    \item $\ket{\sigma_1}=\ket{\Psi^+} = \frac{1}{\sqrt{2}} (\ket{01} + \ket{10})=(\mathbb{I}_2\otimes \sigma_1)\ket{\Phi^+}$ indicates a Pauli-X (\(\sigma_1\)) error.
    \item $\ket{\sigma_2}=\ket{\Psi^-} = \frac{1}{\sqrt{2}} (\ket{01} - \ket{10})=-i(\mathbb{I}_2\otimes \sigma_2)\ket{\Phi^+}$ indicates a Pauli-Y (\(\sigma_2\)) error.
    \item $\ket{\sigma_3}=\ket{\Phi^-} = \frac{1}{\sqrt{2}} (\ket{00} - \ket{11})=(\mathbb{I}_2\otimes \sigma_3)\ket{\Phi^+}$ indicates a Pauli-Z  (\(\sigma_3\)) error.
\end{itemize}

A an interesting feature of this protocol is that the summary remains true regardless of which qubit in each pair is subjected to Pauli errors. Such a strategy can be generalized to measure the $n$-qubit quantum channel $\mathcal{P}$ using $n$ pairs of maximally entangled qubits for $n\geq 2$ thanks to the fact that 
\begin{equation}
    \ket{\Phi_n}=\ket{\Phi^+}^{\otimes n}.
\end{equation}
From each pair, we pick one qubit to form a set. Then, we have two sets of $n$ qubits. Subsequently, one set passes through the channel $\mathcal{P}$. Bell measurements detect errors on each qubit and its corresponding pairs. For the simplest case $n=2$, the circuit in Figure \ref{fig: Bell measurement detecting error n=2} shows how the method is employed.

\begin{figure}[!h]
    \centering
\begin{quantikz}
    \lstick{\ket{0}} & \gate{H} & \ctrl{1} & \qw & \ctrl{1} & \gate{H} & \meter{} & \qw \\
    \lstick{\ket{0}} & \qw      & \targ{}  & \gate[wires=2]{\mathcal{P}} & \targ{} & \qw & \meter{} & \qw \\
    \lstick{\ket{0}} & \gate{H} & \ctrl{1} & \qw & \ctrl{1} & \gate{H} & \meter{} & \qw \\
    \lstick{\ket{0}} & \qw      & \targ{}  & \qw & \targ{} & \qw & \meter{} & \qw \\
\end{quantikz}
    \caption{The Pauli channel is applied on the second and third qubits. Then, the errors on these two qubits are detected by Bell measurements on the initial pairs.}
    \label{fig: Bell measurement detecting error n=2}
\end{figure}

Intuitively, if the tuple of errors $(\sigma_{i_1},\sigma_{i_2},...,\sigma_{i_n})$ is the output of the measurements on $n$ qubit pairs, we get a sample of the distribution $P$ with the value $i=(i_1,i_2,...,i_n)\in \{0,1,2,3\}^n$. Then, one measurement experiment exploiting one use of $\mathcal{P}$ results in one sample, providing an upper bound for the query complexity of the sampling task. We formalize this direct correspondence in the following lemma.
\begin{lemma}
     If Bell measurements are performed on $n$ pairs of qubits using the above strategy, 
the resulting tuple of errors $(\sigma_{i_1}, \sigma_{i_2}, \ldots, \sigma_{i_n})$ 
forms a sample corresponding to the outcome $i = (i_1, i_2, \ldots, i_n) \in \{0,1,2,3\}^n$ 
drawn from the distribution $P$. 
     \label{lemma sample}
\end{lemma}

\textit{Proof.} We denote the post-measurement state corresponding to the tuple of error $\tau_i=(\sigma_{i_1},\sigma_{i_2},...,\sigma_{i_n})$ by $\ket{\tau_i}=\ket{\sigma_{i_1}}\ket{\sigma_{i_2}}...\ket{\sigma_{i_n}}$. Considering the image of the Pauli channel under Choi isomorphism with the remark that $\ket{\sigma_{i_k}}=(\mathbb{I}_2\otimes \sigma_{i_k})\ket{\Phi^+}$ and $\ket{\Phi_n}\bra{\Phi_n}=(\ket{\Phi^+}\bra{\Phi^+})^{\otimes n}$, we have
\begin{align}
    (\mathbb{I}_{A}\otimes\mathcal{P})(\ket{\Phi_n}\bra{\Phi_n})&=\sum_{i\in \{0,1,2,3\}^n}P(i)\,(\mathbb{I}_{A}\otimes\tau_i)\ket{\Phi_n}\bra{\Phi_n}(\mathbb{I}_{A}\otimes\tau_i^\dagger)\\
    &=\sum_{i\in \{0,1,2,3\}^n}P(i)\,\bigotimes_{k=1}^n (\mathbb{I}_2\otimes \sigma_{i_k})\ket{\Phi^+}\bra{\Phi^+}(\mathbb{I}_2\otimes \sigma^\dagger_{i_k})\\
    &=\sum_{i\in \{0,1,2,3\}^n}P(i)\,\bigotimes_{k=1}^n \ket{\sigma_{i_k}}\bra{\sigma_{i_k}}\\
    &=\sum_{i\in \{0,1,2,3\}^n}P(i)\ket{\tau_i}\bra{\tau_i}.
\end{align}
So with probability $P(i)$, the Bell measurements detect the tuple error $\tau_i$.
\newline

\textit{ Proof of \textbf{Theorem \ref{theorem: main bisimulation}}}. We start from the first point: the Pauli channel $\mathcal{P}$ can be simulated utilizing the random variable $P$. Consider a quantum system consisting of $n$ qubits initially in the state $\rho$, on which we aim to simulate the action of the Pauli channel $\mathcal{P}$, given access to the random variable $P$. This process is straightforward. A sample $i=({i_1},{i_2},...,{i_n})$ is drawn from the distribution $P$ and then the corresponding Pauli operation $\tau_i=(\sigma_{i_1},\sigma_{i_2},...,\sigma_{i_n})$ is applied to the $n$-qubit state. The outcome will be the quantum state $\tau_i\rho\tau_i^\dagger$, with probability $P(i)$, i.e. the quantum state is presented by the density operator
\begin{equation}
    \mathcal{P}(\rho)= \sum_{i\in\{0,1,2,3\}^n}P(i)\,\tau_{i}\rho\tau_{i}^\dagger.
\end{equation}

We now consider the second point: obtaining one individual sample from $P$ necessitates exactly one application of the channel $\mathcal{P}$. The lower bound is evident because, given the Pauli channel $\mathcal{P}$, to draw one sample from the distribution $P$, the quantum circuit must query $\mathcal{P}$ at least once. Lemma \ref{lemma sample} confirms that one query is enough. The upper bound and the lower bound match. Therefore, Theorem \ref{theorem: main bisimulation} is proven $\square$.

\subsection{Optimality}
Let $\mathcal{A}$ denote the collection of all algorithms that sample from $P$ using Pauli channel queries. For $a \in \mathcal{A}$, the algorithm applies the Pauli channel $M_a$ times and produces $N_a$ samples from $P$. After Theorem \ref{theorem: main bisimulation}, an interesting question arises: is there any algorithm $a \in \mathcal{A}$ so that $N_a>M_a$? 

The natural conjecture is that no such algorithm exists. Our insight is that the amount of useful information extracted from $M_a$ channel applications cannot exceed the information contained in $M_a$ random variables, which are independent and identically distributed (i.i.d.) according to $P$. Theorem~\ref{theorem: main bisimulation} confirms the equivalence between the information contained in the random variable $P$ and the Pauli channel $\mathcal{P}$. Before presenting a rigorous proof, our intuitive discussion, following simple logic, will suggest that $M_a$ channel applications can provide at most $M_a$ samples of $P$, emphasizing the insight of this bisimulation relation.

Considering $a \in \mathcal{A}$, Theorem~\ref{theorem: main bisimulation} implies that $M_a$ channel applications can be simulated by $M_a$ i.i.d. copies of the random variable $P$. In other words, the $M_a$ channel applications in algorithm $a$ can be replaced by a random variable $X \sim P^{\otimes M_a}$, which produces an output $Y \sim P^{\otimes N_a}$. This results in a modified version of algorithm $a$, and we do not distinguish between the simulated version and the original. We denote the simulation efficiency of $a$ as:
$$
r_a = \frac{N_a}{M_a}.
$$

\textit{Observation.}
Given only $N$ finite samples from a discrete probability distribution $P$, no method can produce an estimate $\hat{P}$ that approximates $P$ arbitrarily well with high probability. Equivalently, with high probability the distance $\ell_p(\hat{P}, P)$ is lower-bounded by some error $\epsilon(N) > 0$, regardless of the learning algorithm employed \cite{waggoner2015lpdistance}.

Suppose, for contradiction, that there exists a simulation algorithm with unbounded simulation efficiency. This would enable the generation of infinitely many samples from a finite initial sample set. A learning algorithm, as described in \cite{waggoner2015lpdistance}, could then exploit these simulated samples to violate the lower bound $\epsilon(N)$. This yields a contradiction. Hence, the simulation efficiency $r_a$ must be bounded for all algorithms $a \in \mathcal{A}$. In fact, we will further show that $r_a \leq 1$ for all $a \in \mathcal{A}$.

% Now assume that there exists a simulation algorithm with unbounded simulation efficiency. Such an algorithm would allow us to produce an infinite number of samples from a finite initial sample set. A learning algorithm, as described in \cite{waggoner2015lpdistance}, could then use these output samples to violate the lower bound $\epsilon(N)$. This leads to a contradiction. Therefore, simulation efficiency $r_a$ must be bounded for all algorithms $a\in \mathcal{A}$. We will go further to see that $r_a\leq 1$ for all $a\in \mathcal{A}$. 

Assume that there exists $a \in \mathcal{A}$ such that $r_a > 1$, i.e., $N_a > M_a$. We describe a family of algorithms $\{b_k\}_{k \in \mathbb{N}} \subset \mathcal{A}$. Algorithm $b_k$ takes $M_a$ copies of $P$ and runs $a$ to obtain $N_a$ copies of $P$. Then, it recycles $M_a$ out of the $N_a$ output copies as input for another execution of algorithm $a$. Each time algorithm $a$ is executed, we gain $N_a - M_a$ additional copies of $P$. Using this procedure, algorithm $b_k$, after $k$ recycles, produces $M_a + k(N_a - M_a)$ copies of $P$ starting from $M_a$ initial copies. The simulation efficiency of $b_k$ is given by:
$$
r_{b_k} = 1 + k \frac{N_a - M_a}{M_a}.
$$

This simulation efficiency grows arbitrarily large as $k$ increases, which contradicts the fact that simulation efficiency is bounded. Therefore, the assumption that $r_a > 1$ is incorrect. Hence, the answer to the initial question is that $r_a \leq 1$ for all $a \in \mathcal{A}$. We state our finding in the lemma below and technically prove it with no-cloning principle \cite{no-cloning-theorem}.
\enlargethispage{1\baselineskip}
\begin{lemma}[Tight Bounds of The Sampling Task] 
Obtaining $N>0$ samples of $P$ requires exactly $M=N$ channel applications.
\label{lemma: tight bound of sampling}
\end{lemma}

\textit{Proof.}
\begin{itemize}
    \item \textit{Upper bound.} The upper bound $M\leq N$ is achieved by the algorithm in Lemma \ref{lemma sample}.

    \item \textit{Lower bound.} In the discussion of the simulation efficiency of sampling algorithms, we primarily used reasoning based on classical information processing. We will now reaffirm our finding using the no-cloning principle. Assume that there exists $a\in \mathcal{A}$ such that $r_a >1$. Then, in the corresponding family $\{b_k\}_{k \in \mathbb{N}} \subset \mathcal{A}$ described above, there exists $b_\kappa$ such that $r_{b_\kappa}\geq 2$. This means that from $M_a$ i.i.d. samples of $P$, we can produce $2M_a$ i.i.d. samples of $P$. We note that any classical distribution can be written in a density matrix form and in our quantum computing framework, any operation we may perform on a quantum state must be modeled by a CPTP map. Let us consider the density matrix $\rho$ that represents the classical state of $M_a$ i.i.d. samples of $P$. Then, algorithm $b_\kappa$ induces a CPTP map $\mathcal{E}_{b_\kappa}$ such that for all classical state $\rho$ that may represent $P$,

    \begin{equation}
        \mathcal{E}_{b_\kappa}(\rho \otimes\ket{int}\bra{int}) = \rho \otimes \rho,
    \end{equation}
    where $\ket{int}\bra{int}$ is the initial state of an ancillary system, which is used to contain information copied from $\rho$. One can easily find that it is a contradiction. Indeed, considering two classical states $\rho_1$ and $\rho_2$, we have that $\rho_{12}=\frac{1}{2}(\rho_1+\rho_2)$ is also a classical state. Algorithm $b_{\kappa}$ is supposed to ensure that
    \begin{equation}
    \begin{cases}
        \mathcal{E}_{b_\kappa}(\rho_{1} \otimes\ket{int}\bra{int}) = \rho_{1} \otimes \rho_{1}\\
        \mathcal{E}_{b_\kappa}(\rho_{2} \otimes\ket{int}\bra{int}) = \rho_{2} \otimes \rho_{2}\\
        \mathcal{E}_{b_\kappa}(\rho_{12} \otimes\ket{int}\bra{int}) = \rho_{12} \otimes \rho_{12}.
    \end{cases}
    \end{equation}
    However in general, if $\rho_1 \neq \rho_2$, these three equalities violate the linearity of CPTP map $\mathcal{E}_{b_{\kappa}}$. This contradiction proves that no such algorithm $b_\kappa$ can exist, which invalidates our initial assumption that $r_a > 1$. Therefore, for any algorithm $a \in \mathcal{A}$, the simulation efficiency $r_a = N_a/M_a \leq 1$. This implies $M_a \geq N_a$, establishing the lower bound.
\end{itemize}

\textit{Proof of \textbf{Theorem \ref{theorem: adaptivity is not useful}}.}  Lemma \ref{lemma: tight bound of sampling} directly leads to the proof of Theorem \ref{theorem: adaptivity is not useful}. Specifically, given the Pauli channel $\mathcal{P}$, in order to prove that adaptivity offers no advantage in sampling from the distribution $P$, we established two points:
\begin{enumerate}
    \item Obtaining $N$ samples of $P$ requires $M = \Theta(N)$ channel applications.
    \item There exists a non-adaptive algorithm that achieves this query complexity. $\square$
\end{enumerate}

To elaborate further on this inference: Consider the general strategy shown in Figure \ref{fig: adaptive strategies} in the context of Theorem \ref{theorem: main bisimulation}, the redundancy of adaptivity is evident. Theorem \ref{theorem: main bisimulation} indicates that just one application of the gate $\mathcal{P}$ suffices to produce a sample from the distribution $P$. After this initial use, the entire system may undergo some possible quantum operations. Subsequently, it is subjected to quantum measurement. Any subsequent steps to interpret the measurement outcome are purely classical and do not involve further quantum operations on the system. The possible quantum operations before the measurement cannot be adaptive because no measurement outcome was available.

\section{Learning and Testing Pauli Channels}
Revisiting the exploration of learning and testing discrete distributions based on $\ell_p$ norms reveals a longstanding research interest. Waggoner has solidified the definitive bounds for the complexities of learning and uniformity testing \cite{waggoner2015lpdistance}. Leveraging his main results allows us to infer the learning complexity, and the white noise testing for Pauli channels  determining if $P$ represents a completely random variable of size $2^n$. The logic is very simple. As discussed above, Theorem \ref{theorem: main bisimulation} and Lemma \ref{lemma: tight bound of sampling} identify the number of samples required to estimate any properties of $P$ with the necessary number of queries to $\mathcal{P}$. Corollaries \ref{corollary tight bound for learning} and \ref{corollary testing white noise} are the consequences of the results in \cite{waggoner2015lpdistance} and the translation between the two kinds of complexities.

\subsection{Learning Algorithm}

We now consider the classical processing part of learning channel $\mathcal{P}$. The algorithm proposed in the reference \cite{waggoner2015lpdistance} is employed to learn the distribution $P$. The procedure for drawing samples has been described in the previous section.

\begin{algorithm}
\caption{Learning the discrete distribution $P$}
\begin{algorithmic}[1]
\State \text{Input:} \text{channel } $\mathcal{P}$; number of qubits $n$ ; distance parameter $p\in [ 1,\infty ] $   and $ \epsilon,   \delta \in (0,1) $
\State \text{Choose} \( N \) sufficient for \( p \), \( n \), \( \varepsilon \), \( \delta \) according to proven upper bounds in Lemma \ref{lemme exact upper bound learning}.
\State \text{Draw} \( N \) samples.
\State \text{Let} \( X_i \) be the number of samples drawn of each coordinate \( i \in \{0, 1,2,3\}^n \).
\State \text{Compute} each \( \hat{P}_i = \frac{X_i}{N} \).
\State \text{Output} \( \hat{P} \).
\end{algorithmic}
\label{alg: learner}
\end{algorithm}
We have the exact sufficient numbers of samples to learn the discrete distribution $P$ using Algorithm \ref{alg: learner} according to Theorem 5.2 in \cite{waggoner2015lpdistance}, applied in the context of distribution size of $4^n$, as written in Lemma \ref{lemme exact upper bound learning}. In addition, Theorem 5.4 in \cite{waggoner2015lpdistance} shows the lower bounds up to constant factors of learning discrete distributions in $\ell_p$, presented in Lemma \ref{lower bound in Wagooner l_p}.

\begin{lemma}[Exact Upper Bounds on Learning Sample Complexity]
To learn $P$ within error $\epsilon \in (0,1)$ with respect to the $\ell_p$ distance and with success probability at least $1-\delta$, 
it suffices to run Algorithm~\ref{alg: learner} with the following number of samples:
\begin{equation}
N_{O} = \frac{1}{\delta} \begin{cases} 
4^{\frac{n(2-p)}{p}}\epsilon^{-2} & \text{if }  1 \leq p \leq 2\hspace{0.1cm}\text{ and } \epsilon \leq 2\cdot 4^{-\frac{n(p-1)}{p}}\\
\frac{1}{4} \left(\frac{2}{\epsilon}\right)^{\frac{p}{p-1}} &  \text{if }  1 \leq p \leq 2\hspace{0.1cm}\text{ and } \epsilon \geq 2\cdot 4^{-\frac{n(p-1)}{p}}\\
\epsilon^{-2} & \text{if } 2\leq p\leq \infty.
\end{cases}    
\end{equation}
\label{lemme exact upper bound learning}
\end{lemma}
\begin{lemma}[Lower Bounds on Learning Sample Complexity] To learn $P$ within error $\epsilon \in (0,1)$ with respect to the $\ell_p$ distance and with constant success probability, 
the required number of samples is at least
\begin{equation}
N_{\Omega} =  
\begin{cases}

\Omega\left( 4^{\frac{n(2-p)}{p}}\epsilon^{-2}\right) & \text{if } 1 \leq p \leq 2 \text{ and } \epsilon \leq 4^{-\frac{n(p-1)}{p}} \\
\Omega\left( (\frac{1}{\epsilon})^{\frac{p}{p-1}} \right) & \text{if } 1 \leq p \leq 2 \text{ and } \epsilon \geq 4^{-\frac{n(p-1)}{p}} \\
\Omega\left( {\epsilon^{-2}} \right) & \text{if } 2 \leq p \leq \infty

\end{cases}  
\end{equation}
    \label{lower bound in Wagooner l_p}
\end{lemma}

The lower bounds and upper bounds in Lemma \ref{lemme exact upper bound learning} and \ref{lower bound in Wagooner l_p} match up to constant factors. We conclude the tight bounds for learning Pauli channel $\mathcal{P}$ in Corollary \ref{corollary tight bound for learning}.

\textit{Proof of \textbf{Corollary \ref{corollary tight bound for learning}}.} The translation from distribution sample complexities to channel query complexities is automatic thanks to Theorem \ref{theorem: main bisimulation} and \ref{theorem: adaptivity is not useful}. Except the tight bound $\Theta(4^n)$ in the regime where $1\leq p\leq 2$ and $4^{-\frac{n(p-1)}{p}} \leq \epsilon \leq 2\cdot 4^{-\frac{n(p-1)}{p}}$, the derivation of the other values of $N$ directly follows from the matching, up to constant factors, between the bounds in Lemma \ref{lemme exact upper bound learning} and \ref{lower bound in Wagooner l_p}. Considering this distinct regime, given the fact that $\epsilon = \Theta(4^{-\frac{n(p-1)}{p}})$, the lower bound and the upper bound converge to the same order, i.e. $(\frac{1}{\epsilon})^{\frac{p}{p-1}}=\Theta(4^n)$ and $4^{\frac{n(2-p)}{p}}\epsilon^{-2}=\Theta(4^n)$ $\square$.

At the critical value $p=2$, all the tight bounds in the corollary converge to $\Theta(1/\epsilon^{2})$. Interestingly, for all $p\geq 2$, and specifically for the $\ell_\infty$ norm, the required number of channel applications remains the same at $\Theta(1/\epsilon^{2})$. When $p=1$, we consistently find ourselves within the regime $\epsilon \leq 2^{-\frac{2n(p-1)}{p}}= 1$, leading to $N=\Theta(4^n/\epsilon^{2})$. Remarkably, the complexities presented in the corollary are optimal up to only constant factors, effectively covering all possible scenarios for general $\ell_p$ norms.

The algorithm employing maximally entangled qubit pairs and Bell measurements demonstrates significant quantum advantages as shown by the comparisons in Table \ref{tab: compare learning bounds}. For the $\ell_\infty$ distance, it refines strategies from previous entanglement-free settings, reducing their complexities from $\Theta(2^n/ \epsilon^{2})$ in \cite{LiangJiang2023tightbound} and $O(1/\epsilon^2)\log(n/\epsilon)$ in \cite{flammia2021pauli_populationrecovery} to $\Theta(1/\epsilon^{2})$. The bounds for the case of learning in the distance $\ell_1$ also benefit from quantum advantage, improved from $O(n2^{3n}/\epsilon^2)$ to $\Theta(4^n/\epsilon^2)$, and tighter than the previous one using entanglement $\Tilde{O}(4^n/\epsilon^2)$. Moreover, for the $\ell_2$ distance, the algorithm reduces the upper bound from $O(n2^n/\epsilon^2)$ to merely $\Theta(1/\epsilon^2)$, eliminating the dependency on the number of qubits. Thus, the principal contribution of this part is the establishment of tight query complexity bounds for learning Pauli channels across any $\ell_p$ distance. Furthermore, the effectiveness of our non-adaptive algorithm further confirms that adaptivity does not provide any additional benefit in learning the Pauli channel when entanglement resources are deployed.

\subsection{White Noise Testing}

% We now consider the problem of testing if the Pauli channel is a white noise source, i.e., $P$ is a uniform distribution over $4^n$ elements of $\{0,1,2,3\}^n$, denoted by $U_{4^n}$. More formally, the goal is to distinguish between the two hypotheses
% \begin{equation*}
%     H_0: P = U_{4^n}\quad \text{ and } \quad H_1: \ell_p(P,U_{4^n})>\epsilon
% \end{equation*}
% with high probability, where the deviation from uniformity is measured by the $\ell_p$-distance with a threshold of $\epsilon$.

We now consider the problem of testing whether the Pauli channel is a white noise source, 
that is, whether $P$ is the uniform distribution $U_{4^n}$ over the $4^n$ elements of $\{0,1,2,3\}^n$. 
More formally, the task is to distinguish between the two hypotheses
\begin{equation*}
    H_0: P = U_{4^n} 
    \qquad \text{and} \qquad 
    H_1: \ell_p(P,U_{4^n}) > \epsilon
\end{equation*}
with high probability, where deviation from uniformity is measured in the $\ell_p$ distance exceeding $\epsilon$.

% The query complexity for this task is derived by translating the classical sample complexity stated in Lemma \ref{lemma uniformity testing classical}, which is a result from \cite{waggoner2015lpdistance}. By substituting the domain size $k=4^n$  into the classical bounds from Lemma \ref{lemma uniformity testing classical}, we directly obtain the results presented in Corollary \ref{corollary testing white noise}.
The query complexity for this task follows by translating the classical sample complexity stated in Lemma~\ref{lemma uniformity testing classical}, originally from \cite{waggoner2015lpdistance}. Substituting the domain size $k=4^n$ into the classical bounds immediately yields the results presented in Corollary~\ref{corollary testing white noise}.

% \begin{lemma}
%     Let $P$ be a distribution over the domain \([k] = \{1, 2, \dots, k\}\) and $U_k$ be the uniform distribution over $[k]$. For $\epsilon \in (0,1)$ the necessary and sufficient sample complexities for the task of distinguishing between the null hypothesis $H_0: P = U_{k}$ and the alternative $H_1: \ell_p(P,U_{k})>\epsilon$ with constant success probability are given in Table \ref{tab:Uniformity test bounds for classical}.
%     \label{lemma uniformity testing classical}
% \end{lemma}

\begin{lemma}[Classical Uniformity Testing Bounds]
Let $P$ be a distribution over the domain $[k] = \{1,2,\dots,k\}$, and let $U_k$ denote the uniform distribution over $[k]$. 
For $\epsilon \in (0,1)$, the necessary and sufficient sample complexities for distinguishing between the hypotheses
\[
H_0: P = U_k 
\qquad \text{and} \qquad 
H_1: \ell_p(P,U_k) > \epsilon
\]
with constant success probability are summarized in Table~\ref{tab:Uniformity test bounds for classical}.
\label{lemma uniformity testing classical}
\end{lemma}

{
\renewcommand{\arraystretch}{1.5} % Adjusts the height of each row
\setlength{\tabcolsep}{20pt} % Adjusts the space between columns
\begin{table}[!h]
\centering
\begin{tabular}{|c|c|c|c|}
\hline
\textbf{p value} & \textbf{Regime} & \textbf{Necessary} & \textbf{Sufficient} \\
\hline
\multirow{2}{*}{\(1 \leq p \leq 2\)} & \( \epsilon \geq k^{-\frac{(p-1)}{p}} \) & \( \epsilon^{-\frac{p}{2(p-1)}} \) & \( \epsilon^{-\frac{p}{2(p-1)}}\) \\
\cline{2-4}
& \( \epsilon < k^{-\frac{(p-1)}{p}} \) & \( k^{\frac{(4-3p)}{2p}}\epsilon^{-2} \) & \( k^{\frac{(4-3p)}{2p}}\epsilon^{-2} \) \\
\hline
\multirow{3}{*}{\(2 < p < \infty\)} & \( \Theta\left( \frac{\log k}{k} \right) \geq \epsilon \) & \( \frac{\log k}{k\epsilon^2} \) & \(\frac{1}{\sqrt{k}\epsilon^2} \) \\
\cline{2-4}
& \( \Theta\left( \frac{\log k}{k} \right) < \epsilon \leq \sqrt{k} \) & \( \frac{1}{\epsilon} \) & \( \frac{1}{\sqrt{k}\epsilon^2} \) \\
\cline{2-4}
& \( \epsilon > \sqrt{k} \) & \( \frac{1}{\epsilon} \) & \( \frac{1}{\epsilon} \) \\
\hline
\multirow{2}{*}{\( p = \infty \)} & \( \Theta\left( \frac{\log k}{k} \right) \geq \epsilon \) & \( \frac{\log k}{k\epsilon^2} \) & \( \frac{\log k}{k\epsilon^2}\) \\
\cline{2-4}
& \( \Theta\left( \frac{\log k}{k} \right) < \epsilon \) & \( \frac{1}{\epsilon} \) & \( \frac{1}{\epsilon} \) \\
\hline
\end{tabular}
\caption{Classical sample complexity for uniformity testing. The bounds indicate the necessary and sufficient number of samples to distinguish a distribution $P$ over $k$ elements from the uniform distribution $U_k$, with respect to the $\ell_p$-distance $\epsilon$.}
\label{tab:Uniformity test bounds for classical}
\end{table}
}

\subsection{Estimating Noise Level}
We begin our discussion on estimating the noise level of Pauli channels by introducing two key characteristic quantities.

\begin{definition}[Shannon Entropy]
The {Shannon entropy} of a distribution \( P \) over the domain \([k] = \{1, 2, \dots, k\}\) is defined as
\[
H(P) = -\sum_{i = 1}^{k} P(i) \log_2 P(i),
\]
where, by convention, \( 0 \log_2 0 = 0 \).
\end{definition}

\begin{definition}[Support Size]
The {support size} of a distribution \( P \) over the domain \([k] = \{1, 2, \dots, k\}\), denoted by \( S(P) \), is defined as the number of elements \( i \in [k] \) such that \( P(i) \neq 0 \).
\end{definition}

Entropy is a fundamental measure of uncertainty in a probability distribution. A higher entropy value indicates greater unpredictability. In this work, we focus on the Shannon entropy, which quantifies the expected amount of information contained in a random variable.

In the context of the Pauli channel, the Shannon entropy of the distribution \( P \) over the domain \( \{0,1,2,3\}^n \) ranges from 0 to \( 2n \) bits. This provides insight into the diversity and randomness of errors affecting the \( n \)-qubit system, enabling a characterization of the channel's noise level. A natural question arises: even if we can estimate the Pauli error probabilities with high precision, does the resulting estimate \( \hat{P} \) yield an accurate approximation of the entropy?

Fortunately, the error in entropy estimation can be bounded using the error in estimating the underlying distribution. Specifically, if the total variation distance between \( \hat{P} \) and \( P \) satisfies \( TV(\hat{P}, P) = \epsilon < \frac{1}{2} \), the Fannes–Audenaert inequality \cite{bound-entropy-errors} ensures that:
\begin{equation}
    |H(\hat{P}) - H(P)| \leq \epsilon \log_2(4^n - 1) + h_{\text{bin}}(\epsilon),
    \label{eq:Fannes-Audenaert-inequality}
\end{equation}
where \( h_{\text{bin}}(\epsilon) = -\epsilon \log_2 \epsilon - (1-\epsilon)\log_2 (1-\epsilon) \) is the binary entropy function.

\textit{Example}. Consider two Pauli distributions \( P \) and \( \hat{P} \) over \( n = 10 \) qubits:
\[
P = \left\{ 0.99, \frac{0.01}{4^{10}-1}, \ldots, \frac{0.01}{4^{10}-1} \right\}, \quad
\hat{P} = \left\{ 0.9, \frac{0.1}{4^{10}-1}, \ldots, \frac{0.1}{4^{10}-1} \right\}.
\]
In this case:
\[
TV(P, \hat{P}) = \epsilon = 0.09, \quad h_{\text{bin}}(\epsilon) \approx 0.436,
\]
\[
H(P) \approx 0.281, \quad H(\hat{P}) \approx 2.469.
\]
The difference between the two entropy values is approximately \( 2.188 \), and the theoretical upper bound from Inequality \ref{eq:Fannes-Audenaert-inequality} is evaluated to about \( 2.236 \). While the bound is fairly tight for small systems, it seems to become increasingly loose as \( n \) grows. This motivates the need for improved estimators that remain accurate in high-dimensional regimes.

Another important characteristic of the Pauli channel is its potential sparsity: not all error \( \tau_i \in \{0,1,2,3\}^n \) necessarily occur with non-zero probability. Moreover, errors may be concentrated in specific regions of the quantum system or limited to certain error types (e.g., bit flips or phase flips). Given the exponential growth of the distribution size, \( 4^n \), obtaining a large enough sample to observe this sparsity directly would require significant resources.

However, if we can accurately estimate the actual support size \( S(P) \) using only a small number of samples, this would complement the entropy estimate and provide a fuller picture of the channel's noise structure. Together, entropy and support size serve as two foundational indicators of the Pauli channel’s behavior.

Breakthrough work by Valiant and Valiant \cite{VV11, VV17} proposes a method to estimate such distributional properties—including both entropy and support size, using sublinear sample complexity. We adopt their estimator, referred to as \emph{``Estimate Unseen''}, and formalize its performance in the lemma below.

\begin{lemma}[Estimate Unseen]
Let \( P \) be a distribution over \([k] = \{1,2,\dots,k\} \), and suppose \( k \in \mathbb{N} \) is sufficiently large. There exist positive constants $\alpha$ and $\gamma$ such that for any \( \epsilon \in (0,1) \), the ``Estimate Unseen'' algorithm, using $\frac{\gamma}{\epsilon^2} \cdot \frac{k}{\log k}$ samples from \( P \), returns with probability at least \( 1 - e^{-k^{\alpha}} \):
\begin{itemize}
    \item An estimate \( \hat{H}(P) \) satisfying \( | \hat{H}(P) - H(P) | < \epsilon \),
    \item An estimate \( \hat{S}(P) \) satisfying \( \frac{| \hat{S}(P) - S(P) |}{k} < \epsilon \), provided that \( P(i) \notin (0, \frac{1}{k}) \) for all \( i \).
\end{itemize}
Furthermore, the sample complexity \( \Theta \left( \frac{1}{\epsilon^2}\cdot\frac{k}{\log k} \right) \) is optimal up to constant factors for this task.
\label{lemma Estimate Unseen}
\end{lemma}

This algorithm functions as a classical information-theoretic tool for analyzing the noise level in the Pauli channel \( \mathcal{P} \). We apply Lemma \ref{lemma Estimate Unseen} to the probability distribution \( P \) that characterizes the Pauli channel $\mathcal{P}$, using the Bell measurement strategy once again to draw samples from $P$ to obtain the formal restatement for Corollary \ref{corollary Estimate the noise level} in the following corollary. This result builds on the complexity translation framework developed in Theorem \ref{theorem: main bisimulation} and Theorem \ref{theorem: adaptivity is not useful}.

\begin{Cor}[Complexity of Noise Level Estimation]
There exists an algorithm satisfying the following property: for \( n\in \mathbb{N} \) sufficiently large, there exist positive constants $\alpha$ and $\gamma$ such that for any \( \epsilon \in (0,1) \), using \( \frac{\gamma}{\epsilon^2} \cdot \frac{4^n}{n} \) queries to the Pauli channel \( \mathcal{P} \), 
\vspace{1mm} the algorithm returns with probability at least \( 1 - e^{-4^{n\alpha}} \):
\begin{itemize}
    \item An estimate \( \hat{H}(P) \) satisfying \( | \hat{H}(P) - H(P) | < \epsilon \),
    \item An estimate \( \hat{S}(P) \) satisfying \( \frac{| \hat{S}(P) - S(P) |}{4^n} < \epsilon \), provided that \( P(i) \notin (0, \frac{1}{4^n}) \) for all \( i \).
\end{itemize}
Moreover, the query complexity \( \Theta \left( \frac{1}{\epsilon^2}\cdot\frac{4^n}{n } \right) \) is optimal up to constant factors.

\label{corollary Estimate the noise level formally}
\end{Cor}

\subsection{Estimating Diamond Distance between Two Pauli Channels}
In this part, we investigate the task of distinguishing between two Pauli channels using the {diamond distance}. Consider a quantum system composed of \( n \) qubits, represented by the Hilbert space \( A \), along with an ancillary system described by a finite-dimensional Hilbert space \( B \). The Pauli channels of interest, denoted by \( \mathcal{P}_1 \) and \( \mathcal{P}_2 \), are modeled as CPTP maps acting on \( \mathcal{L}(A) \), the space of linear operators on \( A \). Meanwhile, \( \mathcal{I}_B \) denotes the identity super-operator on \( \mathcal{L}(B) \).

\begin{definition}[Diamond Distance]
Let \( \mathcal{P}_1\) and \( \mathcal{P}_2\) be two CPTP maps over $\mathcal{L}(A)$. The {diamond distance} between \( \mathcal{P}_1 \) and \( \mathcal{P}_2 \) is defined as
\begin{equation}
    \|\mathcal{P}_1 - \mathcal{P}_2\|_\diamond = \max_{\chi \geq 0,\ \|\chi\|_1 = 1} \left\| \left( \mathcal{I}_B \otimes (\mathcal{P}_1 - \mathcal{P}_2) \right) \chi \right\|_1,
\end{equation}
where \( \chi \) is a density operator on \( B \otimes A \), and \( \|\cdot\|_1 \) denotes the trace norm.
\end{definition}

This measure quantifies the maximum possible deviation between the actions of the two quantum channels, even in the presence of entanglement between the system \( A \) and an arbitrary ancillary system \( B \). Specifically, it captures the difference between two quantum channels even when their action are only on part of an arbitrarily large system. This property makes the diamond distance especially relevant for evaluating the fidelity of quantum operations, where seemingly small discrepancies between an ideal channel and its physical implementation can lead to significant changes in computational outcomes. Consequently, the diamond norm serves as a fundamental metric for verifying the precision and robustness of quantum information processing tasks.

In the context of Pauli channels, where two CPTP maps \( \mathcal{P}_1 \) and \( \mathcal{P}_2 \) are defined by probability distributions \( P_1 \) and \( P_2 \), respectively, an insightful result shown in \cite{PhysRevA.85.042311_diamond_distance_total_variance} establishes that the diamond distance between these two maps is precisely twice the total variation distance (equivalently, the \( \ell_1 \)-distance) between \( P_1 \) and \( P_2 \). Formally, this relationship is given by
\begin{equation}
    \|\mathcal{P}_1 - \mathcal{P}_2\|_\diamond = \sum_{i \in \{0,1,2,3\}^n} |P_1(i) - P_2(i)| = \ell_1(P_1, P_2).
\end{equation}

According to Theorem~1, achieving an \( \epsilon \)-accurate approximation in \( \ell_1 \)-distance for the distributions \( P_1 \) and \( P_2 \), via empirical estimates \( \hat{P}_1 \) and \( \hat{P}_2 \), requires a sample complexity of \( \Theta(4^n / \epsilon^2) \). This implies that approximating the diamond distance between two Pauli channels up to an additive error of \( 2\epsilon \) can be accomplished using \( O(4^n / \epsilon^2) \) queries to the channels. Specifically, the triangle inequality yields
\begin{equation}
    |\ell_1(P_1, P_2) - \ell_1(\hat{P}_1, \hat{P}_2)| \leq \ell_1(P_1, \hat{P}_1) + \ell_1(P_2, \hat{P}_2) < 2\epsilon.
\end{equation}

However, the question regarding the optimality of this approach arises. By exploiting the work of Valiant et al. in \cite{VV11,VV17} and our Bell measurement strategy, we find that the actual number of applications of the Pauli channels needed to estimate $\|\mathcal{P}_1-\mathcal{P}_2\|_\diamond$ effectively reduces to $\Theta(\frac{4^n}{n\epsilon^2})$, indicating a more efficient estimation process. To achieve this improvement, we utilize the \emph{distance estimator} for the \( \ell_1 \)-distance introduced in \cite{VV17}, formalized in the following lemma.

\begin{lemma}
Let \( P_1 \) and \( P_2 \) be probability distributions over the domain \([k] = \{1, 2, \dots, k\}\), where \( k \in \mathbb{N} \) is sufficiently large. Then there exist positive constants $\alpha$ and $\gamma$ such that for any \( \epsilon \in (0,1) \), the \emph{distance estimator}, using \( \frac{\gamma}{\epsilon^2} \cdot \frac{k}{\log k} \) independent samples from each of \( P_1 \) and \( P_2 \), returns, with probability at least \( 1 - e^{-k^{\alpha}} \), an estimate \( \hat{d} \) of \( \ell_1(P_1, P_2) \) satisfying
\begin{equation}
    |\hat{d} - \ell_1(P_1, P_2)| < \epsilon.
\end{equation}
Furthermore, the sample complexity \( \Theta \left( \frac{1}{\epsilon^2} \cdot \frac{k}{\log k} \right) \) is optimal up to constant factors for this task.
\label{lemma estimate diamond}

\end{lemma}

By applying the same complexity translation technique as in Theorem~\ref{theorem: main bisimulation} and Theorem~\ref{theorem: adaptivity is not useful}, this result directly leads to the following corollary tailored to Pauli channels.

\begin{Cor}[Complexity of Diamond Distance Estimation]
There exists an algorithm with the following property: for sufficiently large \( n \), there exist positive constants \( \alpha \text{ and } \gamma \) such that for any \( \epsilon \in (0,1) \), the algorithm uses \vspace{1mm}\( \frac{\gamma}{\epsilon^2} \cdot \frac{4^n}{n} \) queries to each of the Pauli channels \( \mathcal{P}_1 \) and \(\mathcal{P}_2  \), and with probability at least \( 1 - e^{-4^{n\alpha}} \), returns an estimate \( \hat{d} \) of \( \ell_1(P_1, P_2) \) satisfying
\begin{equation}
    |\hat{d} - \ell_1(P_1, P_2)| < \epsilon.
\end{equation}
Moreover, the query complexity \( \Theta \left( \frac{1}{\epsilon^2}\cdot\frac{4^n}{n } \right) \) is optimal up to constant factors for this task.
\label{Cor estimate diamond formally}
\end{Cor}

\section{Conclusion}

Our work has addressed the complexities involved in learning and testing Pauli channels, establishing lower and upper bounds across all $\ell_p$ norms, most of which are tight. Building on foundational work in discrete probability distributions and employing novel algorithms such as ``Estimate-Unseen'', we have provided deeper insights into quantum noise characterization in Pauli channels. Our exploration highlights the cost of accurately estimating critical properties of the error distribution, such as its Shannon entropy, support size, and the diamond distance to between channels. Furthermore, our findings show that adaptivity provides no advantage when maximally entangled qubit pairs and Bell measurements are available. This work contributes to the theoretical framework for quantum information processing and offers practical insights for developing more efficient quantum computing systems. Future research may extend these methodologies to other quantum channel models, opening new directions for quantum error correction and noise mitigation.
\enlargethispage{1\baselineskip}
\bibliographystyle{unsrtnat}
\bibliography{pauli}
% \enlargethispage{1\baselineskip}
\end{document}